\documentclass{article}
\usepackage{spconf,amsmath,graphicx}
\usepackage{graphicx, subfig}
\usepackage{amssymb}
\usepackage{multirow}
\usepackage[cmyk]{xcolor}
\usepackage{fontawesome}
\usepackage{pifont}
\usepackage{makecell}

\title{Continuous interaction  with a smart speaker via low-dimensional embeddings of dynamic hand pose}
\name{Songpei Xu, Chaitanya Kaul, Xuri Ge, Roderick Murray-Smith
\thanks{Thanks to Moodagent for partial funding of S.X. RM-S \& C.K. are partially funded by EPSRC projects EP/R018634/1 and EP/MO1326X/1.}}
\address{School of Computing Science, University of Glasgow, UK}
\begin{document}
\ninept
\maketitle
\begin{abstract}
This paper presents a new continuous interaction strategy with visual feedback of hand pose and mid-air gesture recognition and control for a smart music speaker, which utilizes only 2 video frames to recognize gestures.  

Frame-based hand pose features from MediaPipe Hands,  containing 21 landmarks, are embedded into a 2 dimensional pose space by an autoencoder.
The corresponding space for interaction with the music content is created by embedding high-dimensional music track profiles to a compatible two-dimensional embedding. A PointNet-based model is then applied to classify gestures which are used to control the device interaction or explore music spaces. By jointly optimising the autoencoder with the classifier, we manage to learn a more useful embedding space for discriminating gestures. 

We demonstrate the functionality of the system with experienced users selecting different musical moods by varying their hand pose.

\end{abstract}
\section{Introduction}
\label{sec:intro}

Mid-air gesture recognition and control has recently attracted increasing research attention in multimedia  applications. 
Early works such as Kinect in the Kitchen \cite{panger2012kinect} has explored mid-air gestural control and feedback using a Kinect in cooking scenarios, where common devices have limited displays and touch is less appropriate when cooking. 
Recently, many studies \cite{shakeri2017novel,qian2020aladdin} have proposed mid-air interaction methods for driving based on mid-air gesture recognition and control, which can prevent driver distraction caused by conventional physical handling or touch. 
However, conventional gesture interaction methods require a physical device as support and focus on solving physical controls and interactions, such as simple music control, device selection, \textit{etc.} 
Hence, in recent years deep learning based gesture recognition methods \cite{ur2021dynamic, ahmed2022radar} have been studied to enhance mid-air gesture control and interaction for many applications. 
For instance, \cite{ur2021dynamic} proposed a deep learning architecture based on the combination of a 3D Convolutional Neural Network (3D-CNN) and a Long Short-Term Memory (LSTM) network, which takes advantages of spatial-temporal information from 30-frame video sequences. 
Though they achieve significant improvements, these methods remain unsatisfactory due to long time sequences dependencies and high-dimensional feature inputs of models. 
In addition, the interpretability of the user gesture interaction process is not addressed by most current methods, which is widely regarded as a critical component in real-world gesture control. 

\begin{figure*}
  \centering
  \includegraphics[width=1\textwidth]{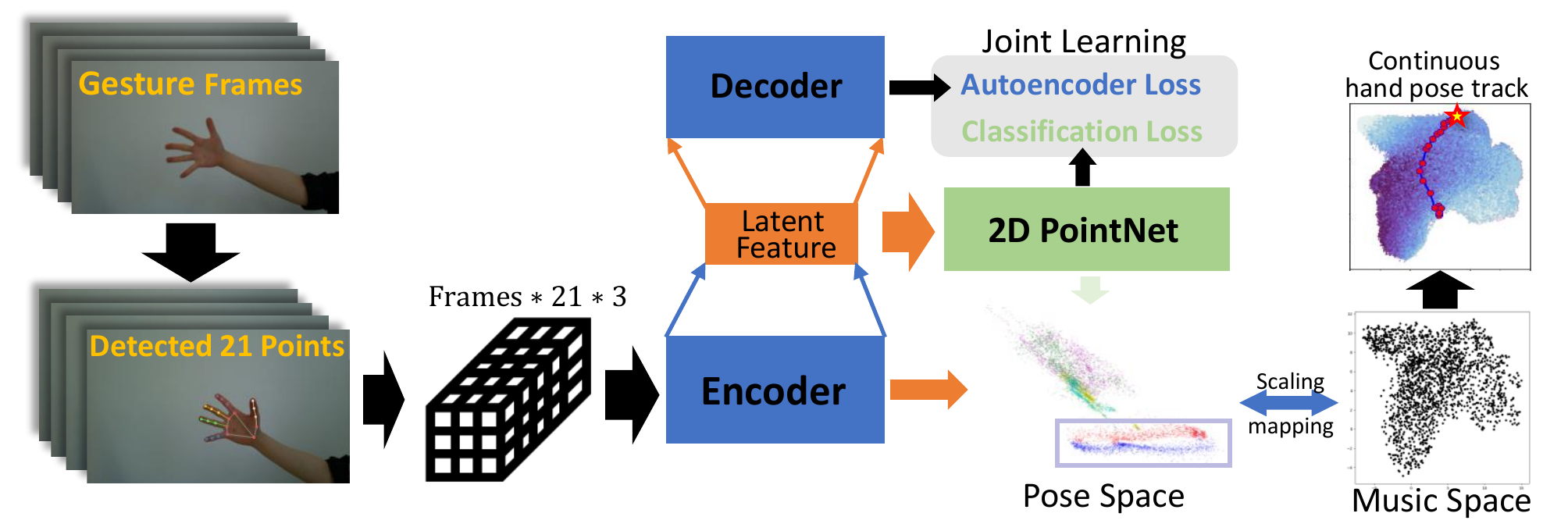} 
  \caption{Pipeline of mid-air gesture control} 
  \label{fig:pipeline} 
\end{figure*}

To deal with the mentioned problems, we propose a straightforward solutions to combine gesture recognition with low-dimensional embeddings, which uses low-dimensional embeddings to reduce the dimension of features of multiple frames and visualize the gestures.
Low-dimensional feature embeddings \cite{zebari2020comprehensive, gisbrecht2015parametric, mcinnes2018umap} 
 are very popular for real-time tasks, where gesture recognition and interaction are temporally connected, due to multiple types of sensors used for data collection. 
These embeddings can be linear mappings based on principal component analysis (PCA) \cite{zebari2020comprehensive} and factor analysis (FA) \cite{gisbrecht2015parametric}, and even non-linear mappings based on t-distributed stochastic neighbor embedding (t-SNE) \cite{gisbrecht2015parametric}, uniform manifold approximation and projection (UMAP) \cite{mcinnes2018umap}, and autoencoders -- all creating effective explorations of low-dimensional embeddings. 
However, the linear nature of PCA 
limits the ability to project the features to a lower dimension without losing information.
t-SNE and UMAP can perform non-linear feature dimensionality reduction, but they cannot reduce the additional costs associated with subsequent classification.
In this work, we design a fully-connected autoencoder to reduce the dimension of the detected hand pose features and to learn a better low-dimensional pose space for interaction. 
The autoencoder takes hand pose features as inputs, reduces their dimensionality via an encoder to get low-dimensional latent features, and then reconstructs them using a decoder. 
This is an unsupervised process, and by training using a dataset containing numerous gestures, we can obtain generalization ability and expressiveness. 
Our proposed autoencoder reduces the gestures to an interactive 2D space, which facilitates visualization of the embedded pose  space mapped with the corresponding music space to provide a more intuitive hand pose based exploration.  
The temporal information from video sequences input in the gesture or action recognition model is exploited to further improve the classification accuracy in many studies \cite{herath2017going,ur2021dynamic}. 
However, the dependence of longer frame sequences makes them difficult to be truly interactive in real time. And for gesture recognition and control applied to realistic scenarios, interaction response delays tend to annoy users and reduce perceived usability \cite{anderson2011user, ng2012designing}. 
In this paper, we propose a simple PointNet-based classification network to recognize the predefined discrete gestures and continuous hand poses by fewer frames (2 frames) in low-dimensional inputs (2 dimensions) from an autoencoder. 
Discrete gestures mean that feedback is obtained after the full gesture has been triggered, while continuous gestures obtain the real-time feedback while the gesture is in progress. 
Compared to the methods \cite{kim2010wearable,vogiatzidakis2021mid,liao2021smart} that focus only on discrete gesture interaction, our approach handles both discrete and continuous gesture-based interaction scenarios.
Finally, we defined corresponding functions for the different recognized gestures, which include discrete gesture control for music start/stop and continuous hand pose control for real-time musical space exploration, \textit{etc}.
Different from other gestural interaction strategies \cite{lv2014multimodal} and video process methods \cite{neimark2021video}, our proposed pipeline overcomes to some extent the disadvantages of high-dimensional feature input and long sequence dependence, and  implements a continuous hand pose to explore the music space.
Specifically, we map the predicted continuous hand pose to a music space with different properties, where the two-dimensional hand pose space is embedded from the autoencoder encoder. 
The low frame dependency of the gesture recognition stage gives our model the advantage of low latency. 
Visible user interaction based on the autoencoder encoding gives the user more freedom of choice and exploration, which is not exploited in the literature. 

\section{Methodology}
\label{sec:methodology}
Fig. \ref{fig:pipeline} presents a detailed structure of our proposed pipeline for interaction with a smart music speaker. It includes generating a low-dimensional embedding (Section \ref{ssec:dimension reduction}), gesture classification (Section \ref{ssec:Classification}), and Interaction (Section \ref{ssec:Interaction}).

\subsection{Low-dimensional Embedding}
\label{ssec:dimension reduction}
In this work, the main purpose of the low-dimensional embedding is that  the space distribution of low-dimensional hand pose features can be visualized.
In this way, gesture interactions are more understandable and more controllable when interacting with a music space. 
Specifically, MediaPipe Hands \cite{lugaresi2019mediapipe,zhang2020mediapipe}, a real-time hand landmarks detection model, is used to extract 3-dimensional (3D) coordinate information of 21 hand landmarks, which saves a significant model space compared to directly using pixel-level image (image size is $480 \times 620$). 
Then, an autoencoder based on fully connected layers \cite{rusu2022low} is designed to reduce the dimensionality of the 3D coordinates of the 21 hand landmarks. 
The encoder and decoder in autoencoder contain four fully-connected layers respectively, and the layers use LeakyReLU \cite{xu2015empirical} for non-linear activation. 
We set the neuron numbers of 4 fully-connected layers in the encoder are 128, 96, 64 and 2, respectively and the reverse in the decoder.

\subsection{Gesture Classification}
\label{ssec:Classification}  %
We employ a highly efficient and effective PointNet \cite{qi2017pointnet} that directly consumes point information, based on multiple linear layers,  to classify the predefined gestures. 
PointNet-based classification network reduces the model size and training time by using low-dimensional inputs from the autoencoder, as well as assisting the pose space to obtain better distinguishability in the autoencoder by their jointly learning. 
Furthermore, inspired by the popular sequence-based methods \cite{ur2021dynamic, neimark2021video}, we also explore and take advantage of the sequence information from the gesture video. 
Specifically, the frame-based sequence features are encoded by the autoencoder as new inputs of the classification network. 
Then, we get the corresponding predicted gesture categories from the optimized classification network.
In this work, we explore single-, 2- and 8-frame sequence inputs for the PointNet-based classification, and we chose a 2-frame sequence as our final inputs, based on trading-off classification performance with minimizing time delay. 

Our goal is to jointly learn the parameters of the proposed fully-connected layer based autoencoder and PointNet-based classification by minimizing a loss
function over the training set, which employs mean square loss \cite{sara2019image} and cross-entropy loss \cite{zhang2018generalized}, respectively. We employ Adam \cite{kingma2014adam} with 0.001 weight decay and 0.001 learning rate as the optimizer of our joint model.


\subsection{Interaction}
\label{ssec:Interaction}
In this work, we anticipate that users will interact with the smart music speaker through a combination of discrete gestures and continuous hand poses. 
We propose a novel, user-friendly strategy for control interaction and exploration in a visible music space, where a discrete gesture (Pinch) is used for activation and the continuous hand pose (Continuous arm open/close) is used for continuous exploration. 
And other gestures will be used for other interactions in future works. 
Our music data is provided by our industrial partner, Moodagent, including about 55,000 music tracks. Each track is represented by 34 features, including predicted subjective scores for 6 emotion types, 14 genre types, and 14 style types. These features are derived from the track's audio signal using a convolutional neural network to predict human subjective classifications. The music features are embedded by UMAP down to a 2-dimensional music space for human interaction. In this work, we focus on the exploration of a music space with different emotions, including sadness, joy, fear, erotic, anger and tenderness, which the user can interact with via continuous hand pose changes. 
To connect the pose space to the music space, we use a physical mapping, \textit{i.e.}, first scaling the music space and the pose space to the same range and computing the cluster centres for each type of music and then computing the distance to the coordinates of the real-time pose in the two-dimensional pose space. The music category with the closest distance to the coordinates of the real-time pose will be highlighted. 
We mark different emotions in different colours in the music space. The colours range from light to dark, indicating light to heavy emotional expressions of music. 
In this way, users have more freedom and controllability to explore the music space with an entire music database by the continuous mid-air hand pose movement, as shown in Fig. \ref{fig:pipeline}.
In addition, we explore the possibility of using alternative representations, such as quaternion \cite{rieger2004systematic, saxena2009learning}, to make pose spaces more stable.

\section{EXPERIMENTS}
\label{sec:experiments}
\subsection{Dataset}
\label{ssec:dataset}
We investigate and design interactive control gestures that conform to human habits for the smart music speaker, including 'continuous arm open' (the arm and hand away from body), 'continuous arm close' (the arm and hand close to body), index finger drawing 'circle clockwise', drawing 'circle counterclockwise', 'pinch' and 'double-pinch'. 
In fact, since these gestures are in pairs and opposite directions, we collect 6 gestures in total.
Specifically, using the Intel$^\circledR$ RealSense$^\text{TM}$ LiDAR Camera L515 depth camera (frame rate is 30 fps), we collected 25 clip videos for each gesture for 7 volunteers. 
Each video duration varies from 1 to 3 seconds. 
To avoid the influence of the background, we choose a white wall about one meter away from the camera as the background and the collected gestures are 40-50 cm away from the camera.
After that, we extract over 60, 000 frames containing gestures from the collected clip videos. 
In Table \ref{tab:gesture}, we provide the detailed information of our collected gesture dataset.

\begin{table}[t!]
\scriptsize
\begin{center}
\fontsize{8}{10}\selectfont
\renewcommand\tabcolsep{8.0pt}
\caption{Overview information of our collected Gesture frames.} \label{tab:gesture}
\begin{tabular}{cc|cc|c}
\hline
\multicolumn{1}{c|}{No.} & Gesture                 & Train & Test  & Total \\ \hline
\multicolumn{1}{c|}{1}          & Continuous arm open          & 8793  & 2209  & 11002 \\
\multicolumn{1}{c|}{2}          & Continuous arm close         & 9096  & 1711  & 10807 \\
\multicolumn{1}{c|}{3}          & Circle clockwise        & 8647  & 1873  & 10520 \\
\multicolumn{1}{c|}{4}          & Circle counterclockwise & 8517  & 1943  & 10460 \\
\multicolumn{1}{c|}{5}          & Pinch                   & 7588  & 1593  & 9181  \\
\multicolumn{1}{c|}{6}          & Double pinch            & 6742  & 1513  & 8255  \\ \hline
\multicolumn{2}{c|}{Total}                                & 49383 & 10842 & 60225 \\ \hline
\end{tabular}
\end{center}
\vspace{-2.5em}
\end{table}

\subsection{Experimental Results}
In this section, we first present an empirical finding that highlights the effectiveness of a well-selected fully-connected autoencoder in the proposed pipeline, which will get low-dimensional pose spaces for interactions. 
Then the effectiveness of well-designed PointNet-based classifier will be proved. 
Finally, the visualization of mid-air hand pose interactions for a smart music speaker will be given. 

\begin{figure}[t]
  \centering
  \includegraphics[width=1\linewidth]{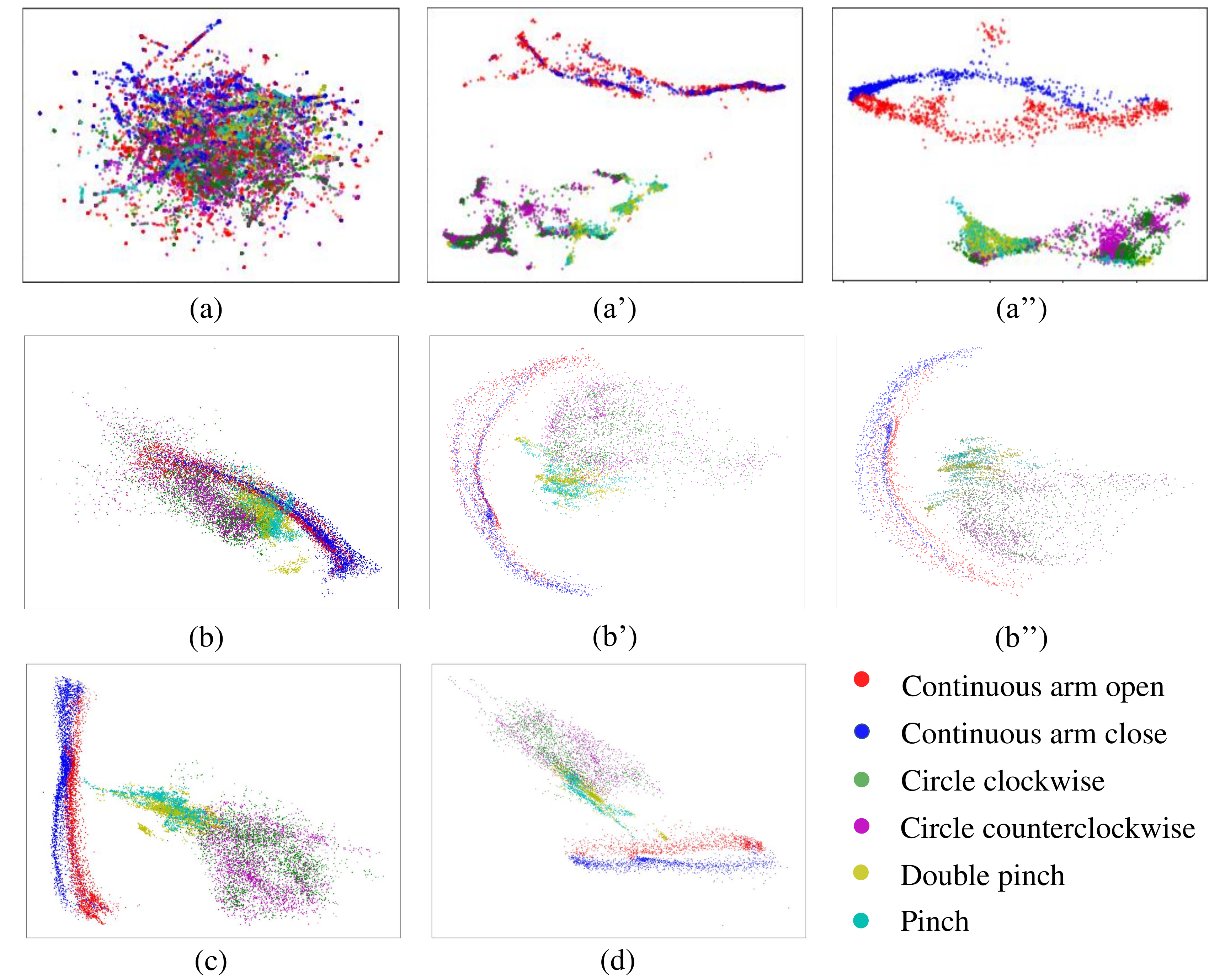} 
  \caption{Visualization of different embedding methods of mid-air gestures. (a), (a') and (a") use single-, 2- and 8-frame gesture sequence inputs of UMAP \cite{mcinnes2018umap}, respectively; (b), (b') and (b'') use single-, 2- and 8-frame gesture sequence inputs of our fully-connected autoencoder without classification; (c) and (d) are joint autoencoder and PointNet-based classifier on single- and 2-frame gesture sequences, respectively. 
} 
  \label{fig:autoencoder} 
\end{figure}

\textbf{The effectiveness of the autoencoder.} 
As shown in Fig. \ref{fig:autoencoder}, the 2D outputs of the different gestures from the encoder in the autoencoder are plotted in a 2D space on the display. 
Compared with the widely used UMAP (indicated by (a), (a') and (a")), which has significant effects on dimensionality reduction \cite{mcinnes2018umap} and clustering \cite{becht2018evaluation}, the proposed fully-connected layer based autoencoder (indicated by (b), (b') and (b")) can better distinguish the distribution of different gestures with lower model complexity.
This allows users to more clearly and intuitively see the positions of different gestures in the pose space and the relationship between different gestures.
When hand pose points in the pose space are sufficiently dispersed, subtle hand pose changes will be clearly tracked. 
Furthermore, we explore the distributional effects of using different frame sequences (2-frame and 8-frame) on the encoding of the proposed fully-connected autoencoder. 
By comparing the visualization results with the first column of Fig. \ref{fig:autoencoder} with single-frame inputs, using time-sequence information can significantly improve inter-class gesture clustering and intra-class dispersion. 
As low latency is very important for user interaction, we focus on 2 frames in our subsequent study, which can avoid long time latency caused by longer sequence dependencies in other methods \cite{ur2021dynamic}. 

\begin{figure}[t]
  \centering
  \includegraphics[width=1\linewidth]{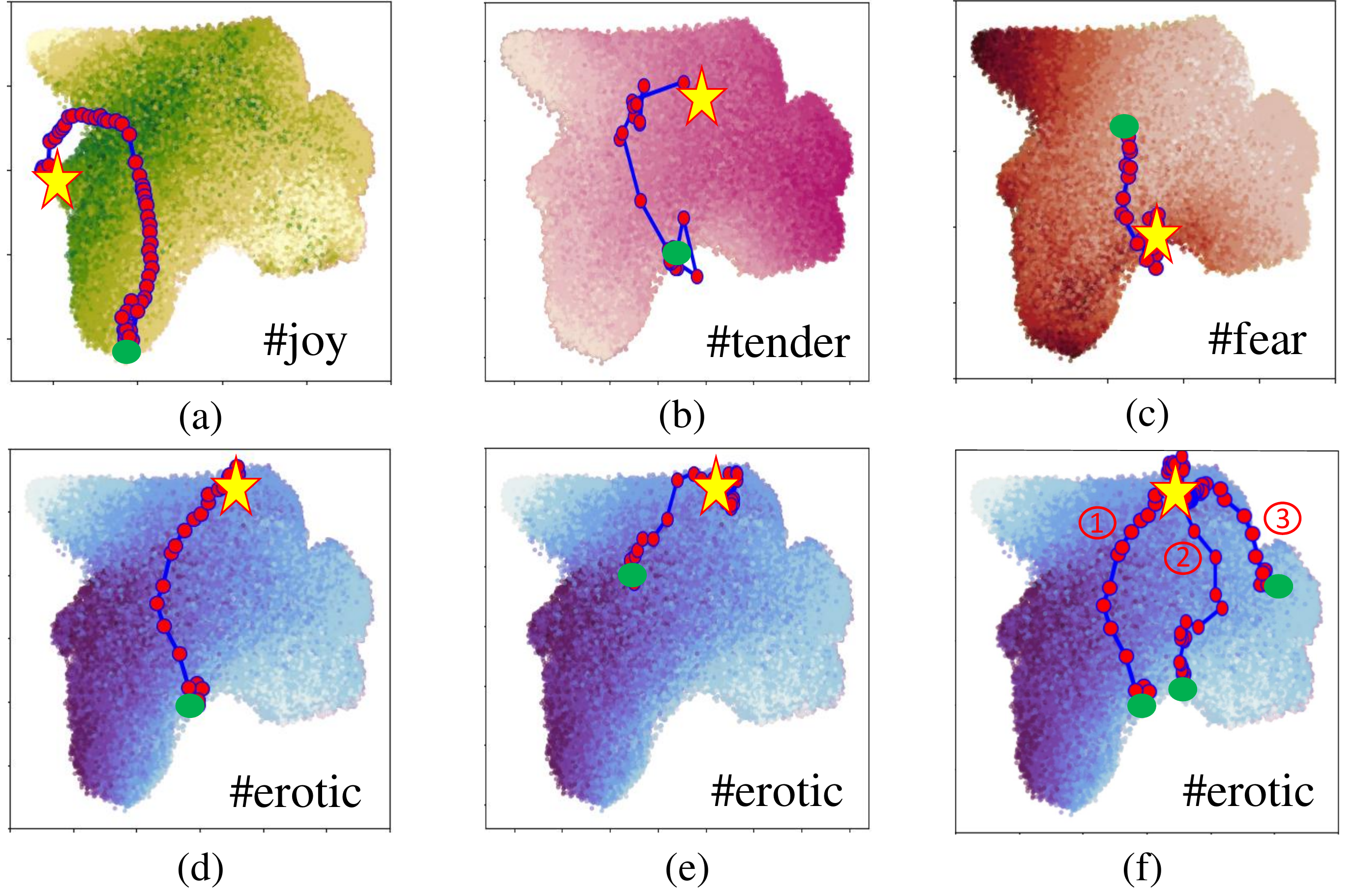} 
  \caption{Visualization of hand-pose-controlled track selection in the music spaces of 55,000 music tracks. The yellow \ding{73} indicates the target music position point, the green \faCircle \ indicates the hand pose starting point, and hand-pose-selected movement tracks are in blue. Different colours of points in the space represent music with different emotions, and the darker the colour, the higher the emotional value.
} 
  \label{fig:music_space} 
\end{figure}
\begin{table}[t]
\scriptsize
\begin{center}
\fontsize{9}{12}\selectfont
\renewcommand\tabcolsep{2.0pt}
\caption{Comparisons of Classification models } \label{tab:cls} 
\label{tab:tab1_class}
\begin{tabular}{c|c|cccc}
\hline
Method                      &   Autoencoder       & Frames & Precision(\%) & mAP(\%) & time (ms) \\ \hline
\multirow{4}{*}{\rotatebox{90}{\makecell[c]{PointNet \\ based}}}     &  $-$ & 1       &  70.6         & 62.1    &  12.6    \\  
                              & $\surd$ & 1      &   71.3        & 61.9    &  2.3    \\
                              &  $\surd$ & 2      &   75.2        & 62.5    &  2.4    \\
                              &  $\surd$ & 8      &   77.2        &  64.9   &   3.0   \\ \hline
\end{tabular}
\end{center}
\end{table}
\textbf{Effects of classification.} 
Different from previous approaches, \textit{e.g.} \cite{rusu2022low}, where only clustering is used for gesture reduction and visualization, in this paper we utilize a PointNet-based classification network to recognize the mid-air gestures for interactions and to further guide the visualization of the distribution of pose space.
As shown in the third row of Fig. \ref{fig:autoencoder}, by joint learning with the classification network, the proposed autoencoder can better distinguish the space distribution of inter- and intra-class mid-air gestures. Although this increases the training time, there is no additional cost in the inference process of gesture feature dimensionality reduction and clustering. 
In addition, in Table \ref{tab:tab1_class}, we provide the detailed classification results for different frame-based gestural sequence features from the autoencoder. 
We also provide inference time for the entire mid-air hand pose interaction process of each sequence.
Compared to the 12.6 ms required by the original PonitNet without autoencoder, the recognition and interactions with autoencoder take 2.3 ms, 2.4 ms and 3 ms on 1-, 2- and 8-frame based sequences, respectively.
It demonstrates that the joint learning of the fully-connected autoencoder and PointNet-based classifier can improve the clustering effect of the autoencoder while also keeping the accuracy of the classification. 
Finally, by balancing classification effectiveness and low latency requirements, we finally choose 2-frame sequences as inputs.

\begin{figure}[t]
  \centering
  \includegraphics[width=1\linewidth]{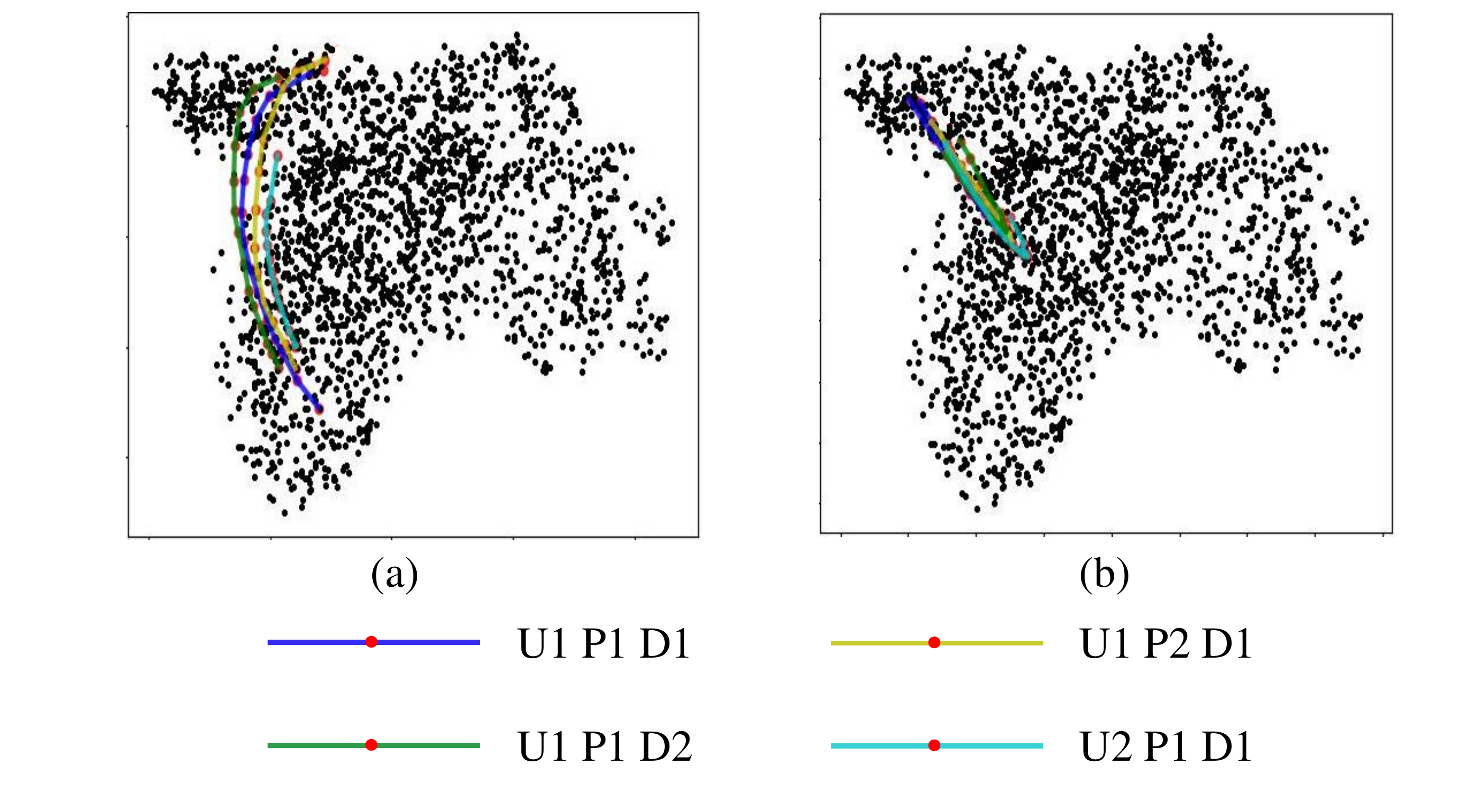} 
  \caption{The figure (a) shows the low-dimensional embedding of 'continuous arm open' when quaternion is not used, and the figure (b) shows the low-dimensional embedding of `continuous arm open' when quaternion is used. In contrast to figure (a), the use of quaternion in figure (b) brings more stability to the model. For greater visibility, the background Music Space using a subset (2,000 music tracks) of the music dataset is shown in black. 
} 
  \label{fig:quaternion} 
\end{figure}


\textbf{Visualization of user interaction and exploration.} 
In order to provide a better experience and understanding when using the smart music player, we  display the selected music position and use the dynamic hand pose continuous control process to map to locations in the music space. 
As shown in the top row of Fig. \ref{fig:music_space}, when we specify different target music positions in music spaces, they can all be reached by continuous movement and exploration of mid-air hand pose. 
For the Fig. \ref{fig:music_space} (d), (e), we measure the time for an experienced user to reach the specified target point when exploring the pose space for two consecutive times, 4.4 s and 2.1 s, respectively. 
This demonstrates that users can learn to explore the music space by continuous dynamic mid-air hand pose control to enhance their understanding of interaction with the music space, and thus reach the goals faster. The three different tracks of hand pose in the Fig. \ref{fig:music_space} (f) proves that users can reach the same target music position with continuous control by different dynamic hand pose, and that the starting position and pose of the hand does not affect the exploration of the target music. 
Notably, the delay of the interaction process of a frame-based gesture sequence (2 frames) is about 2.4 ms, as shown in Table \ref{tab:cls}, including the inference of autoencoder and classifier and drawing. 

In addition, we find that different experienced users with different distances from the camera usually produce different control results on the interaction with the music space for the same hand pose. To have a more stable interaction with the music space, we further explore the use of quaternion \cite{saxena2009learning} to avoid the effect of hand size, hand position in the camera field, and distance from the camera on gesture embedding. 
Fig. \ref{fig:quaternion} compares two experienced users (indicated U1 and U2) of different palm sizes (palm width and palm length) without and with quaternion conversion (as (a) and (b)), respectively, to control the music space at different locations (start position and distance) from the camera using the same gesture. 
Specifically, the palm width and length for U1 are 7 cm and 15 cm respectively, and for U2 are 10 cm and 17 cm, where palm width is the distance from the widest part of the palm and palm length is the distance from the root of the palm to the tip of the middle finger. P1 and P2 indicate different starting positions, with a difference of 20 cm. D1 and D2 indicate different distances from the camera, which are 45 cm and 100 cm, respectively. 
Compared with (a) and (b) in Fig. \ref{fig:quaternion}, it suggests that the use of quaternion effectively reduces the effects of hand size, hand position and distance from the camera on the low-dimensional embedding of the gestures, thus allowing the interactive system to work more stably.

\section{Conclusion}
\label{sec:Conclusion}
In this work, we study the problem of mid-air hand pose control and visible interaction for a smart music speaker and propose a novel pose space encoding and visualization model by a fully-connected autoencoder joined with a PointNet-based gesture classification network. 
Specifically, a new mid-air gesture dataset is collected to train and evaluate the proposed mid-air gesture recognition and control method. 
The proposed autoencoder embeds gestures into low-dimensional spaces suitable for visualisation and interaction, which helps to unify the pose space with the music space for interactions.
In addition, the auxiliary classification network further improves the clustering of gestures in pose space and maintains outstanding classification performance. 
Moreover, the proposed interaction strategy requires only a few gesture frames (2-frame sequence) of input to get a continuous control that the user can explore,
which contributes to the reduction of interaction latency. 

The paper provides an exploratory demonstration of the ability to control and select different areas of the user space via continuous hand pose changes by experienced users. 
\bibliographystyle{IEEEbib}
\bibliography{submission}


\end{document}